\def\apjs{ApJS}                                     
\def\aap{AAP}                                       
\def\lsim{\lower.5ex\hbox{$\; \buildrel < \over \sim \;$}}
\def\gsim{\lower.5ex\hbox{$\; \buildrel > \over \sim \;$}}
\documentclass{aa}
\usepackage{epsfig}

\begin{document}


\title{Spectral analysis of the Galactic e$^{+}$e$^{-}$ annihilation emission}

\author{P.~Jean\inst{1}
         \and J.~Kn\"odlseder\inst{1}
         \and W.~Gillard\inst{1}
         \and N.~Guessoum\inst{2}
	 \and K,~Ferri\`ere\inst{3}
         \and A.~Marcowith\inst{1}
         \and V.~Lonjou\inst{1}
         \and J.P.~Roques\inst{1}}
\institute{ 
$^{1}$ CESR, CNRS/UPS, B.P.~4346, 31028 Toulouse Cedex 4, France \\
$^{2}$ American University of Sharjah, College of Arts \& Sciences, 
Physics Department, PO Box 26666, Sharjah, UAE \\
$^{3}$ LATT, CNRS/OMP, 31000 Toulouse, France \\
}

   \date{Received  ; accepted }

   \authorrunning{Jean et al.}

   \titlerunning{Spectral analysis of the Galactic e$^{+}$e$^{-}$ annihilation emission}


\abstract{   
We present a spectral analysis of the e$^{+}$e$^{-}$ annihilation emission 
from the Galactic Centre region based on the first year of measurements 
made with the spectrometer SPI of the INTEGRAL mission.
We have found that the annihilation spectrum can be modelled by the 
sum of a narrow and a broad 511 keV line plus an ortho-positronium 
continuum. The broad line is detected (significance 3.2$\sigma$) 
with a flux of (0.35 $\pm$ 0.11) $\times$ 10$^{-3}$ photons s$^{-1}$ cm$^{-2}$. 
The measured width of 5.4$\pm$1.2 keV FWHM is in agreement with the
expected broadening of 511 keV photons emitted in the annihilation of 
positroniums that are formed by the charge exchange process of 
slowing down positrons with hydrogen atoms. The flux of the narrow line 
is (0.72 $\pm$ 0.12) $\times$ 10$^{-3}$ photons s$^{-1}$ cm$^{-2}$ and its width 
is 1.3$\pm$0.4 keV FWHM. The measured ortho-positronium continuum 
flux yields a fraction of positronium of (96.7$\pm$2.2)\%.

To derive in what phase of the interstellar medium positrons 
annihilate, we have fitted annihilation models calculated for 
each phase to the data. 
We have found that 49$^{+2}_{-23}$ \% of the annihilation emission 
comes from the warm neutral phase and 51$^{+3}_{-2}$\% from the warm 
ionized phase. While we may not exclude that less than 23\% of the emission 
might come from cold gas, we have constrained the fraction of 
annihilation emission from molecular clouds and hot gas to be less 
than 8\% and 0.5\%, respectively.
  
We have compared our knowledge of the interstellar medium in the 
bulge (size, density, and filling factor of each phase) and the 
propagation of positrons with our results and found that they are in good 
agreement if the sources are diffusively distributed and if the initial 
kinetic energy of positrons is lower than a few MeV. Despite its large 
filling factor, the lack of annihilation emission from the hot gas is
due to its low density, which allows positrons to escape this phase.
 

}

\maketitle


\section{\label{s1}Introduction}

In the quest for the origin of positrons, images of the annihilation line 
emission tell us that positrons annihilate primarily in the bulge of our 
Galaxy (Kn\"odlseder et al. 2005 and references therein). Assuming that 
positrons do not propagate far from their sources, the spatial 
distribution of the annihilation emission should trace the spatial 
distribution of the sources. Under this hypothesis, the observations 
of the spectrometer SPI onboard the INTEGRAL observatory show that the 
sources of the bulk Galactic positrons seem to be associated with the 
old stellar population (Kn\"odlseder et al. 2005). 

In this paper we try to infer, from the spectral characteristics
of the annihilation emission measured by SPI, information on the
particular processes involved in the interaction of Galactic positrons 
with the interstellar medium (ISM). This information should provide 
some clues regarding the origin of Galactic positrons. 
The identification by spectral analysis of the ISM phase in which 
positrons annihilate could enable one to retrieve the type of positron 
sources under particular assumptions of the distance travelled by 
positrons as a function of their initial kinetic energy.
For instance, if positrons propagate a short distance from their 
sources, it is then likely that the positron sources belong to
or are specifically tied to the medium in which positrons annihilate.

The spectral characteristics of the annihilation emission (shape and 
intensity of the line, relative intensity of the ortho-positronium 
continuum) offer valuable information on the physical conditions 
of the ISM where positrons annihilate 
(Guessoum, Ramaty \&\ Lingenfelter 1991; Guessoum, Jean \&\ 
Gillard 2005). Several reports on observations with Ge spectrometers 
suggest a width of the line in the 2--3 keV range (Smith et al. 1993; 
Leventhal et al. 1993; Harris et al. 1998). 
Using only OSSE data, Kinzer et al. (1996) derived a positronium 
fraction of 0.97$\pm$0.03. Measurements with the Ge detector 
TGRS onboard the WIND mission (1995-1997) gives a compatible 
value of 0.94$\pm$0.04 (Harris 1998). From the line width and the 
positronium fraction measurements, Harris et al. (1998) concluded 
that a scenario in which annihilation does not occur either in cold 
molecular clouds or in the hot phase of the ISM is favored. Using 
preliminary SPI data of Jean et al. (2003) and TGRS data of Harris et 
al. (1998), Guessoum et al. (2004) showed that the bulk of the annihilation 
occurs in warm gas. However, they do not exclude that a significant 
fraction of the annihilation may occur in hot gas and in interstellar dust.  
Recently, Churazov et al. (2005) inferred from SPI measurements that 
the spectral parameters of the emission can be explained by positrons 
annihilating in a warm gas or in a combination of warm and cold gases.

In the present work, we include in the spectral analysis the classical 
model of the ISM described by McKee \&\ Ostriker (1977). In this 
model, the ISM consists of molecular clouds, atomic gas in either a 
cold or a warm phase, and ionized gas in either a warm or a hot phase. 
Each phase is characterized by particular physical conditions in 
abundance, temperature, ionization fraction and density. Since the 
annihilation process and the Doppler broadening depend on the target 
properties (H atoms, electrons, velocity...) positrons annihilating 
in a given phase emit a particular spectrum. For instance, positrons 
in a cold medium annihilate mostly by forming positronium in flight, 
whereas the dominant process in a warm ionized medium is radiative 
recombination with free electrons. The spectral characteristics of 
the annihilation in the various ISM phases were first studied by 
Guessoum et al. (1991). They were recently revisited by 
Guessoum, Jean \&\ Gillard (2005) -- herafter GJG05 -- in view of 
the most recent results on positron interaction cross sections with 
H, H$_{2}$ and He as well as a detailed study on the annihilation 
in dust grains.

In the next section, we present the SPI observations and the method used 
to analyse the spectral distribution of the annihilation emission. 
We take into account the SPI spectral response (continuum Compton, 
energy resolution and line deformation due to radiation damage). 
In section 3, we present the results of the spectral analysis. 
We adopt two different approaches consisting of (a) an adjustment 
of simple Gaussian and ortho-positronium laws and (b) a fit of the ISM 
phase fractions using the spectral characteristics of the annihilation 
in each phase calculated by GJG05. This approach differs from that of
Churazov et al. (2005) who fit the temperature and ionized fraction of 
the gas where annihilation occurs with a measured spectrum 
based on $\sim$4.5 $\times$ 10$^{6}$ s duration SPI observations of the 
Galactic Centre region. In section 4, we discuss the implications of these 
results for the origin and physics of Galactic positrons and in section 5, 
we summarize the most important new information. 


\section{\label{s2} Observations and analysis methods}


\subsection{\label{s21} Observations and data preparation}

The data analysed in this work are those of the December 10, 2004 public 
INTEGRAL data release. The data span the IJD epoch 1073.394--1383.573, 
where IJD is the Julian Date minus 2 451 544.5 days. In order to reduce 
systematic uncertainties in the analysis, we exclude observation periods 
with strong instrumental background fluctuations. These background 
variations are generally due to solar flares or the exit and entry of 
the observatory into the Earth's radiation belts. After this cleaning, we 
obtain a total effective exposure time of 15.2 Ms. The exposure 
is rather uniform in the central regions of our Galaxy from where most 
of the annihilation signal is observed (see Kn\"odlseder et~al. 2005).

The spectrum is extracted by model fitting, assuming that the
sky intensity distribution is a $\Delta l \approx$ 8$^{o}$ and 
$\Delta b \approx$ 7$^{o}$ FWHM 2D-Gaussian. 
This distribution is one of the best fitting models derived by 
Kn\"odlseder et~al. (2005) who studied in detail the morphology of 
the annihilation emission. 
This spatial distribution does not 
correspond to the spatial distribution of the Galactic 
diffuse continuum emission, which is expected to be more extended 
in longitude and less extended in latitude. Therefore the intensity of 
the Galactic diffuse continuum emission, calculated in the spectral 
analysis, is overestimated (see section 3.1). However, this systematic 
error does not affect the shape of the annihilation spectrum since the
Galactic continuum emission is fitted by a power-law and, consequently, 
this overestimation factor is the same over the whole energy band.

We perform the analysis in 1 keV 
wide energy bins for a 200~keV wide spectral window covering 
400--600 keV. Each energy bin is adjusted to the data for each 
germanium detector separately, assuming that the count rate is 
due to the sum of the sky contribution and the instrumental 
background. The latter is assumed to be the sum of 3 terms: 
a constant, a component proportional to the rate of saturating 
GeD events (GEDSAT rate) and a component proportional to the 
convolution product of the GEDSAT rate with a 352~day 
exponential decay law (see Eq. 1 in Kn\"odlseder et~al. 2005). 
The last two components are tracers of the short time scale and 
radioactive build up variation of the instrumental background 
at 511 keV, respectively.

\begin{figure}[tb]
\centering
\includegraphics[width=8.8cm,height=6.0cm]{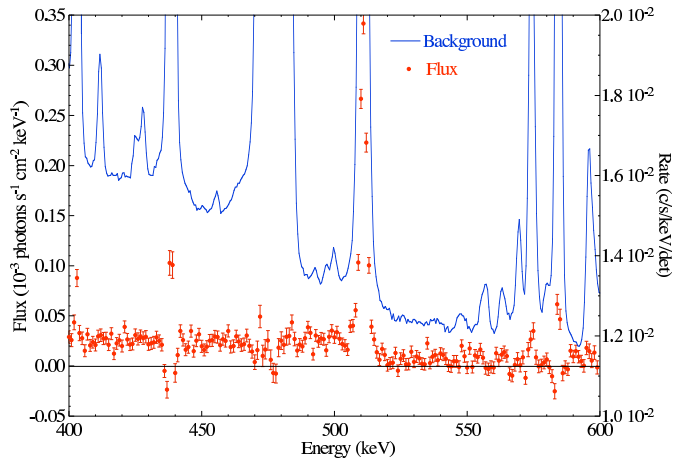}
\caption{Spectrum obtained by model fitting (see text). The
instrumental background spectrum is shown for comparison. \label{fig:bckmf}}
\end{figure}

Figure \ref{fig:bckmf} shows the resulting spectrum. For comparison, 
and to illustrate the quality of the background subtraction, we also 
show the detector averaged count rate spectrum. While overall the
background subtraction is satisfactory, significant residuals 
remain at the location of some of the strong background lines. 
For instance, the 439~keV line, emitted in the decay of the metastable 
state of the $^{69}$Zn isotope is not properly removed. This 
metastable state has a half-life of $\approx$14h and consequently the
temporal variation of this background line differs from the background 
model used in the model fitting. This gives rise to residuals that may 
induce systematic errors in the spectral analysis.

\begin{figure}[tb]
\centering
\includegraphics[width=8.8cm,height=6.0cm]{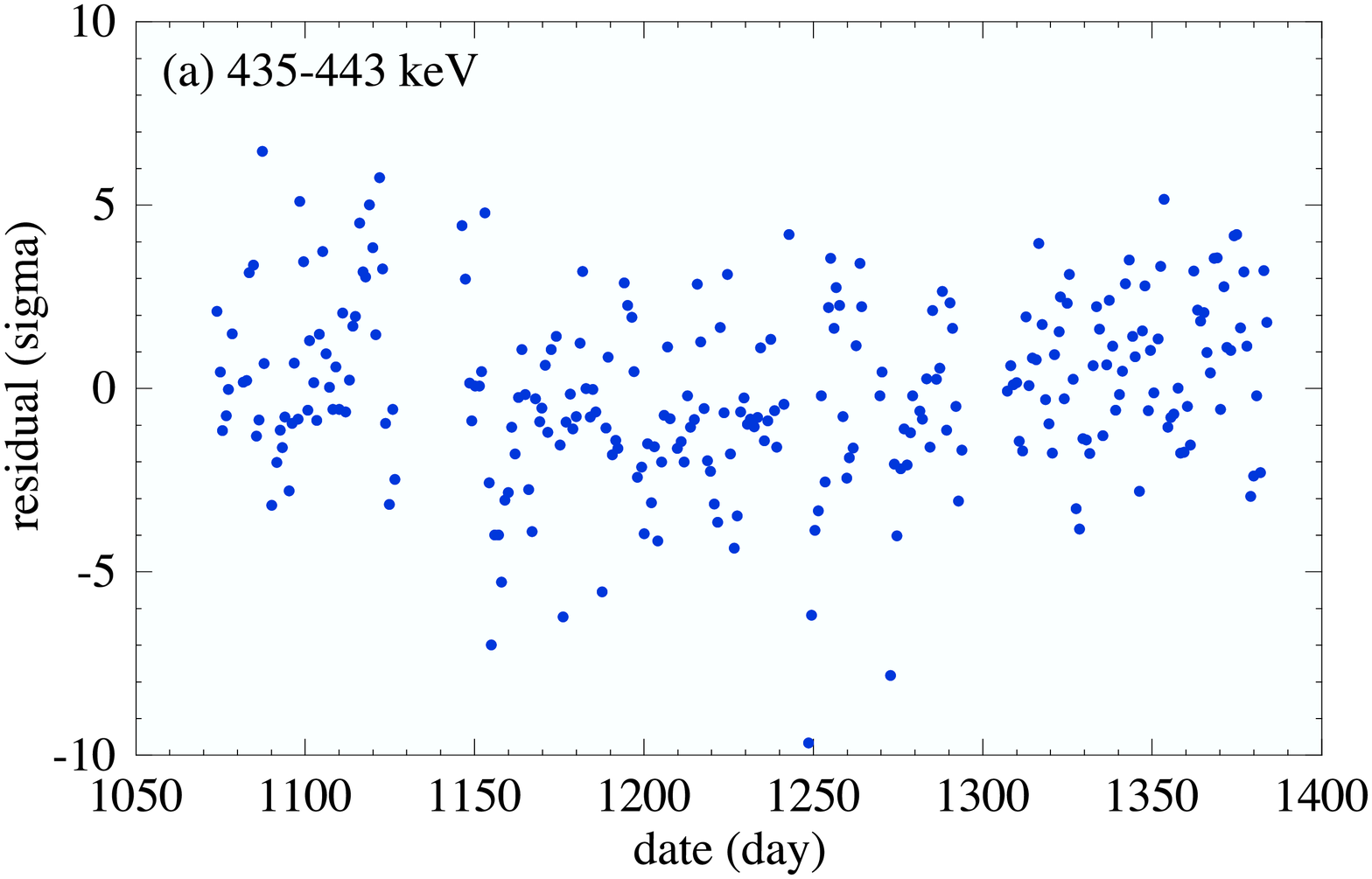}
\includegraphics[width=8.8cm,height=6.0cm]{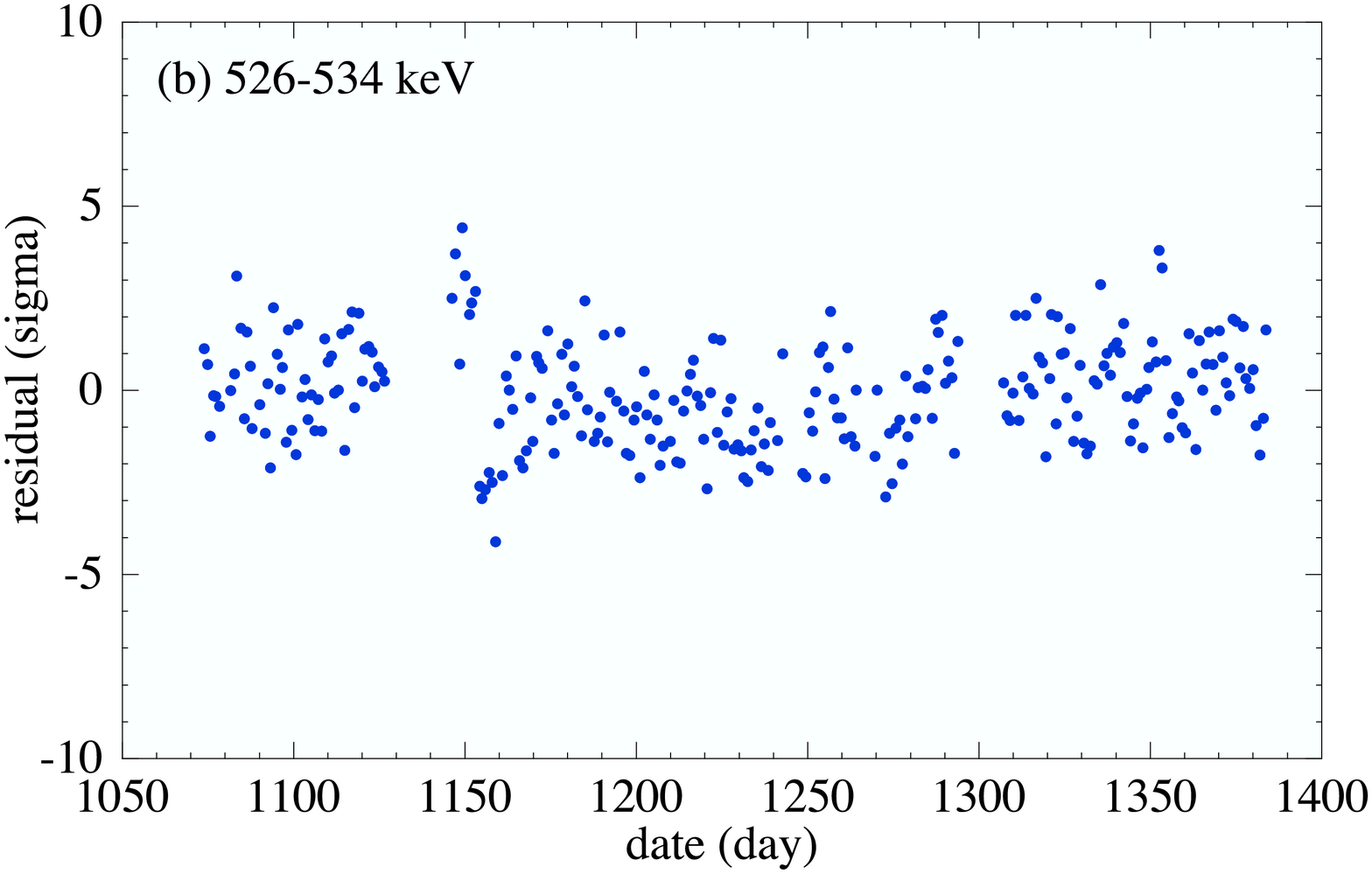}
\caption{Time series of residuals in number of sigma unit obtained by 
subtracting the background plus sky models to the data. Two energy bands 
are shown for comparison: (a) 435--443 keV and (b) 526--534 keV. The 
date are in IJD (see text).
\label{fig:resid}}
\end{figure}

Figure \ref{fig:resid} presents the residuals of the detector averaged 
count rate after subtraction of the background and sky models in 
the 435--443 keV and 526--534 keV bands. The residuals of the 
first band, which contains the 439~keV line, show large fluctuations 
leading to a reduced $\chi^{2}$ of 5.4, while the reduced $\chi^{2}$
is 1.9 for the 526--534 keV band which is free of strong background lines.

\subsection{Spectral analysis methods}

In order to avoid influences of instrumental background lines on the 
spectral analysis, we select the following energy bands, which are relatively 
free of systematic residuals (see previous subsection): 
406--435, 443--467, 480--570, 578--580 and 589--593~keV. They cover 
$\approx$75\%\ of the extracted energy range. This is sufficient (1) 
to quantify accurately the flux of the ortho-positronium continuum and 
(2) to characterize the shape of the annihilation line.

The spectral response of the instrument to a narrow (Dirac like) gamma-ray 
line is an instrumentally broadened Gaussian plus 
a Compton continuum. Due to radiation damage, the shape of the Gaussian is 
deformed with a low energy tail\footnote{High energy 
protons and neutrons impinging on the detectors displace Ge atoms in the 
crystal and then increase the number of hole traps. These traps reduce 
the number of collected holes, leading to an underestimation of the 
energy released by the photon in the detector.}. This warping 
may affect the determination of the physical line width and has to 
be taken into account in the analysis.
A simple way to model radiation damage is to convolve 
a Gaussian  with a decreasing exponential 
function of energy (see Eq. \ref{eq:response}). The energy scale 
$\epsilon_{d}$ of the exponential function quantifies the level of 
degradation and is 
called the ``degradation parameter'' hereafter. The degradation 
increases with time but is regularly removed by the annealing process 
(Roques et al. 2003). Analysis of the shapes of adjacent background lines
yields an energy resolution at 511 keV of 2.0 keV FWHM and a mean 
degradation parameter $\epsilon_{d}$ = 0.3 keV, in agreement with 
previous analyses (Roques et al. 2003; Lonjou et al. 2005).

The Compton continuum shape and level at 511 keV is extracted from the 
IRF (Imaging Response Function) and RMF (Redistribution Matrix File)
available in the SPI data processing database. These reponse 
matrices were generated by Monte-Carlo simulations (Sturner et al. 2003). 
Since the annihilation emission has an extent of $\approx$8$^{o}$ FWHM 
and INTEGRAL scans this emission, we average the IRFs in the 
field-of-view of SPI. 

The spectral response function $R(E)$ is:

\begin{equation}
    R(E) = G(E,\Gamma_{inst}) \otimes e^{-E/\epsilon_{d}} + C(E)
    \label{eq:response}
\end{equation}

\noindent
where $G(E,\Gamma_{inst})$ is a Gaussian with a FWHM of $\Gamma_{inst}$ = 
2.0 keV, $\otimes$ denotes a convolution product and $C(E)$ is the 
Compton continuum function. We then convolve models of the astrophysical 
signal, $S(E,p)$, with this spectral response and fit 
the set of parameters $p$ using the measured spectrum.

We adopt two different approaches to characterize the spectral 
distribution of the annihilation emission:

In the first approach, the ``independent 
model'', we model the spectrum by four independent 
components: two Gaussians $G(E,\Gamma_{i})$ (to model narrow and 
broad 511 keV lines of FWHM $\Gamma_{i}$), the ortho-positronium 
continuum $O(E)$ and a power law to account for Galactic diffuse 
continuum emission. 
The independent model $S_{l}(E)$ is described by:

\begin{eqnarray}
    S_{l}(E) & = & I_{n} \times G(E,\Gamma_{n}) \; + \; I_{b} \times G(E,\Gamma_{b}) \nonumber \\
         & + & I_{3\gamma} \times O(E) \; + \; A_{c} \left( 
         \frac{E}{{\rm 511keV}} \right)^{s}
    \label{eq:splaw}
\end{eqnarray}

where $I_{n}$, $\Gamma_{n}$, $I_{b}$ and $\Gamma_{b}$ are the flux and 
width (FWHM) of the narrow and broad lines, respectively. $I_{3\gamma}$ 
is the flux of the ortho-positronium continuum, which is represented by 
the Ore \&\ Powell (1949) function $O(E)$. $A_{c}$ is the amplitude 
of the Galactic continuum at 511~keV and $s$ is the slope of the power 
law spectrum.

In the second approach, hereafter called the ``ISM model'', 
we adopt the spectral characteristics (line shape and 
ortho-positronium continuum relative flux) for the different 
ISM phases given in the model calculated by GJG05, and with these 
spectral characteristics, we adjust the phase fractions $f_{i}$ 
so as to obtain the best fit to the measured spectrum.
The ``ISM model'' is described by:

\begin{equation}
    S_{ISM}(E) \, = \, I_{e^{+}e^{-}} \times \sum_{i=1}^{5} f_{i} \times 
    S_{i}(E,x_{gr}) \, + \, A_{c} \left( \frac{E}{{\rm 511keV}} \right)^{s}
    \label{eq:spism}
\end{equation}
\noindent
where $S_{i}(E,x_{gr})$ is the normalized spectral distribution 
(in keV$^{-1}$) of the annihilation photons in phase $i$ with 
$i$=$\{$molecular, cold, warm neutral, warm ionized, hot$\}$, 
$I_{e^{+}e^{-}}$ is the annihilation flux (photons s$^{-1}$ 
cm$^{-2}$) and $x_{gr}$ represents the fraction of dust grains ($x_{gr}$ = 1 in 
the standard grain model of GJG05); $x_{gr}$ allows for uncertainties in 
dust abundance and positron-grain reaction rates. Annihilation in dust 
grains is significant only in the hot phase where the standard grain 
model ($x_{gr}$ = 1) yields a FWHM of $\approx$ 2 keV, as opposed to 
a line width of $\approx$ 11 keV in the absence of grains ($x_{gr}$ = 
0). The grain fraction has a negligible effect in the molecular, cold 
and warm neutral phases, and affects the 511 keV line width in the 
warm ionized gas by less than 2\%. The grain fraction is a free 
parameter of the fit, and so too are the phase fractions $f_{i}$, 
subject to the requirement that $\sum_{i=1}^{5} f_{i}$ = 1.

In both cases, the spectral distribution of the Galactic continuum 
emission is assumed to be a power law with amplitude $A_{c}$ at 
511~keV and a fixed slope of $s$ = -1.75 as derived by OSSE measurements 
in this energy range (Kinzer et al. 1999; Kinzer et al. 2000). 
The energy band analysed is not large enough and the SPI exposure 
not sufficient to constrain the slope accurately from the data 
themselves. We also fix the line position at 511 keV since we 
observe a relatively symmetric distribution of the emission around 
the Galactic Centre and do not expect a Doppler shift due to Galactic 
rotation in this region. Moreover, Churazov et al. (2005) did not 
measure a significant shift of the line centroid.

The models (Eqs. \ref{eq:splaw} \&\ \ref{eq:spism}) are convolved 
with the SPI spectral response, $R(E)$ (Eq. \ref{eq:response}). The 
parameters $p$ of each model are fitted by minimizing the $\chi^{2}$. 
The individual errors on the best fit parameters are obtained by 
calculating their confidence intervals for which $\Delta\chi^{2} <$ 1 
(1$\sigma$ uncertainty) with $\Delta\chi^{2} = \chi^{2}(p)-\chi^{2}_{opt}$ 
and $\chi^{2}_{opt}$ the minimum value of $\chi^{2}(p)$. 


\section{Results}


\subsection{\label{s31} Independent model}

\begin{figure}[tb]
\centering
\includegraphics[width=8.8cm,height=6.0cm]{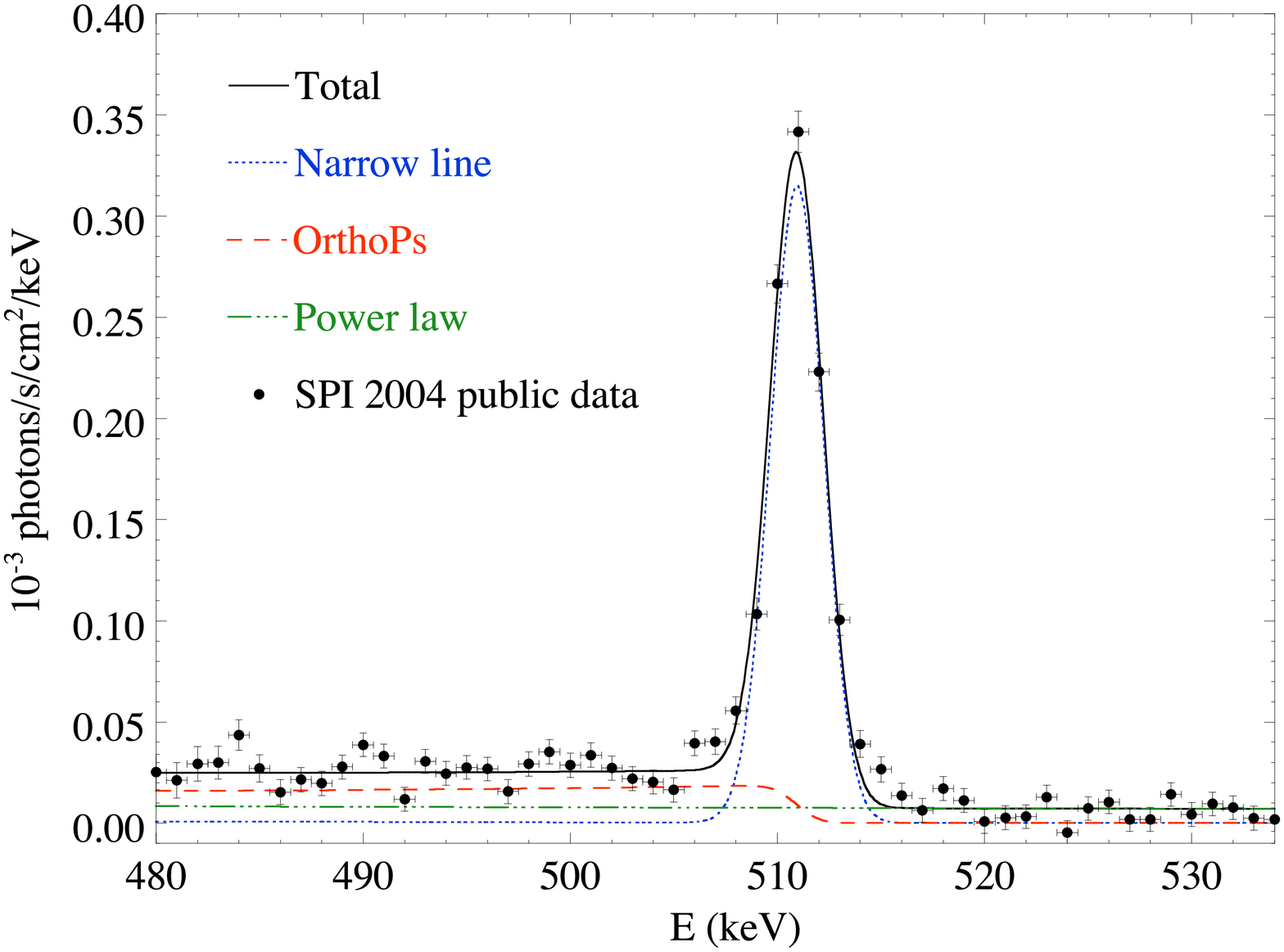}
\caption{Fit of the spectrum measured by SPI with contributions from a 
Gaussian line, an ortho-positronium continuum and a power-law 
Galactic continuum. A single Gaussian does not give a good fit to the 
flux measured in the wings of the line. \label{fig:single}}
\end{figure}

When fitting the line with a single Gaussian, we obtain a 
$\chi^{2}$ of 193.7 for 150 degrees of freedom, a line width of 
(2.2$\pm$0.1) keV, a 511 keV line flux of (1.01$\pm$0.02) $\times$
10$^{-3}$ photons s$^{-1}$ cm$^{-2}$ and an ortho-positronium flux 
of (4.3$\pm$0.3) $\times$ 10$^{-3}$ photons s$^{-1}$ cm$^{-2}$. 
We clearly see a significant excess of counts in the wings 
of the narrow line, suggesting the presence of a broad line (see 
Fig. \ref{fig:single}). This broad line is presumably due to 
the annihilation of the para-positronium state formed in flight. 
We then include the broad line in the analysis and use the model 
presented in Equation \ref{eq:splaw}.

\begin{figure}[tb]
\centering
\includegraphics[width=8.8cm,height=6.0cm]{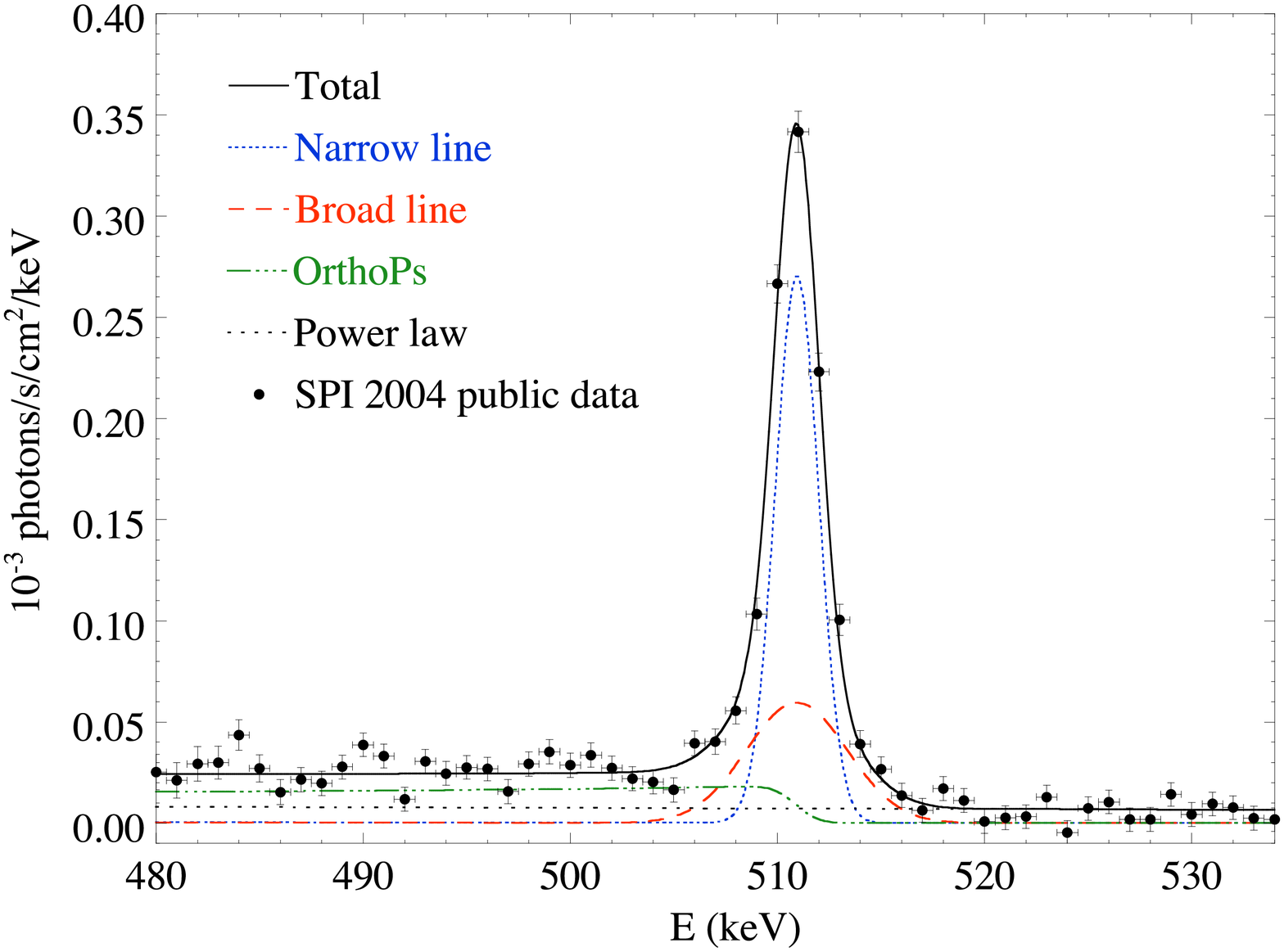}
\caption{Fit of the spectrum measured by SPI with narrow and 
broad Gaussian lines, an ortho-positronium continuum and a power-law 
Galactic continuum (constant slope of -1.75). Note that the asymmetric 
shape of the lines is due to the convolution of the Gaussian with the spectral 
response of SPI (Compton continuum and degradation). \label{fig:spifit1}}
\end{figure}

In this model, the parameters to be adjusted are the widths of the narrow 
and broad lines, their intensities, the ortho-positronium continuum 
flux and the level of the Galactic diffuse continuum. 
Figure~\ref{fig:spifit1} shows the result of the fit. 
We obtain a $\chi^{2}$ value of 171.3 for 148 degrees of 
freedom. The $\chi^{2}$ value is improved by 22.4 with respect to the 
$\chi^{2}$ value obtained when we fit the line with a single Gaussian.

The measured parameters are listed in Table \ref{tab:spifit1}. 
The detection significance of the broad line is 3.2$\sigma$. Its FWHM 
(5.36 $\pm$ 1.22 keV) is in agreement with the width of the 
annihilation line of positronium formed in flight in H (5.8 keV; see GJG05).
The total line (narrow+broad) flux is $I_{2\gamma}$=(1.07$\pm$0.03)$
\times$10$^{-3}$ photons s$^{-1}$ cm$^{-2}$, in agreement with the 
flux derived by Kn\"odlseder et al. (2005) for the 2D Gaussian shaped 
emission profile.
The measured ortho-positronium flux yields a ratio 
I$_{3\gamma}$/I$_{2\gamma}$ of 3.95 $\pm$ 0.32 and consequently 
a positronium fraction of f$_{Ps}$ = 0.967 $\pm$ 0.022, also in agreement 
with previous measurements (Kinzer et al. 1999; Harris et al. 1998; 
Churazov et al. 2005). The statistical uncertainties of 
I$_{2\gamma}$ and I$_{3\gamma}$/I$_{2\gamma}$ are not obtained by combining 
quadratically the uncertainties of I$_{3\gamma}$, I$_{n}$ and I$_{b}$ 
as listed in Table \ref{tab:spifit1}, since these parameters are not 
independent. Instead we fit the parameters of the function:

\begin{eqnarray}
    S_{l}(E) & = & I_{2\gamma} \times [ f_{n} G(E,\Gamma_{n}) + (1-f_{n}) G(E,\Gamma_{b}) \nonumber \\
         & + & R_{3\gamma/2\gamma} O(E) ] \; + \; A_{c} \left( 
         \frac{E}{{\rm 511keV}} \right)^{s}
    \label{eq:splaw2}
\end{eqnarray}
\noindent
where $R_{3\gamma/2\gamma}$ = I$_{3\gamma}$/I$_{2\gamma}$ and f$_{n}$ 
is the fraction of 511~keV flux in the narrow line. 
The width of the narrow line (1.32$\pm$0.35~keV) can be explained 
by thermalized positrons annihilating either in the warm neutral
medium (1.16 keV) or in the warm ionized medium (0.98 keV).  
We also searched for a component with a 11 keV FWHM as expected from 
positrons annihilating in a hot (10$^{6}$ K) interstellar 
grain-free gas. We obtain only an upper limit of 0.36 $\times$ 10$^{-3}$
photons s$^{-1}$ cm$^{-2}$ (2$\sigma$) for such a component, whereupon 
we conclude that annihilation in a hot plasma contributes less than 
$\approx$7$\%$ to the total annihilation flux.

The Galactic continuum intensity at 511~keV ($A_{c}$) is slightly larger 
than the $\approx$ 4--6 $\times$ 10$^{-6}$ photons s$^{-1}$ cm$^{-2}$
keV$^{-1}$ obtained by Kinzer et al. (1999) with OSSE measurements. 
However, these authors extracted this spectral component assuming a 
uniform distribution along the Galactic plane with a latitude width 
of 5$^{o}$ FWHM, while we assume a latitude width of 7$^{o}$ 
FWHM. This can explain the factor $\approx$7/5 discrepancy
between the two estimations.

In order to estimate systematic errors, we first performed the 
analysis with the uncertainty of the spectrum bins increased by a 
factor such that the reduced $\chi^{2}$ is equal to 1. This factor is 
found to be $\approx$1.075, yielding 7.5$\%$ systematic errors. 
Secondly, we quantified the effect on the results of possible uncertainties 
in the fixed parameters (degradation parameter, slope of the continuum). 
The uncertainty in the degradation parameter affects 
neither the continuum intensities nor the total 511 keV flux. Performing 
the analysis with $\epsilon_{d}$ = 0.2 keV and 0.4 keV yields 
differences of less than 1.5$\%$ of the statistical errors in these 
parameter values. Nevertheless, the narrow and broad line widths (and 
fluxes) change by 14$\%$ (and 13$\%$) and 4.5$\%$ (and 13$\%$) of their 
statistical uncertainties, respectively. We adopt these values as 
systematic errors for the parameters of the lines.
Kinzer et al. (1999) did not provide uncertainties in the slope of the 
Galactic Centre continuum spectrum measured by OSSE.
Kinzer et al. (2001) used a slope of -1.65 to adjust the Galactic continuum 
model to the OSSE data. Considering the results presented in Table 3 of 
Kinzer et al. (1999), we can reasonably assume an uncertainty of 
$\pm$0.1 in the slope. This yields a systematic error of 
$\approx$10$\%$ in the intensity of the ortho-positronium continuum 
in our analysis, the other parameters being not significantly affected 
by this change (less than $\approx$1$\%$ of the statistical 
errors). The corresponding systematic uncertainty in the fraction of 
positronium is then $\pm$0.3\%.

We also constructed another spectrum by model fitting (see section \ref{s21}) 
but excluding regions of observations for which the Crab nebula and 
Cyg X-1 are in the field-of-view of SPI, in order to check whether 
these sources produce a bias in the analysis results. The 
statistical error bars of the spectrum are larger due to the smaller 
amount of data in this dataset. The analysis of this spectrum 
provides results that are statistically consistent with those 
presented in Table \ref{tab:spifit1}.

\begin{table}
     \caption{Best-fit values of the free parameters ($\chi^{2}$=171.3 
     for 148 degrees of freedom). $I_{n}$, $\Gamma_{n}$, $I_{b}$ and 
     $\Gamma_{b}$ are the flux and width (FWHM) of the narrow and broad 
     lines, respectively. $I_{3\gamma}$ is the flux of the ortho-positronium 
     continuum and $A_{c}$ is the amplitude of the Galactic continuum at 511~keV. 
     The first set of error bars refers to the 1$\sigma$ 
     statistical errors and the second set to the systematic errors. 
     \label{tab:spifit1}}
     \begin{array}[b]{ll}
\noalign{\smallskip}
\hline
\hline
\noalign{\smallskip}
\mbox{Parameters}  & \mbox{Measured values} \\
\hline
\noalign{\smallskip}
I_{n} \; \mbox{(10$^{-3}$ s$^{-1}$ cm$^{-2}$)} & 0.72 \pm 0.12 \pm 0.02  \\
\noalign{\smallskip}
\Gamma_{n} \; \mbox{(keV)} & 1.32 \pm 0.35 \pm 0.05 \\
\noalign{\smallskip}
I_{b} \; \mbox{(10$^{-3}$ s$^{-1}$ cm$^{-2}$)} & 0.35 \pm 0.11 \pm 0.02 \\
\noalign{\smallskip}
\Gamma_{b} \; \mbox{(keV)} & 5.36 \pm 1.22 \pm 0.06 \\
\noalign{\smallskip}
I_{3\gamma} \; \mbox{(10$^{-3}$ s$^{-1}$ cm$^{-2}$)} & 4.23 \pm 0.32 \pm 0.03 \\
\noalign{\smallskip}
A_{c} \; \mbox{(10$^{-6}$ s$^{-1}$ cm$^{-2}$ keV$^{-1}$)} & 7.17 \pm 0.80 \pm 0.06 \\
\noalign{\smallskip}
  \hline
\end{array}
\end{table}


\subsection{ISM model}

\begin{figure}[tb]
\centering
\includegraphics[width=8.8cm,height=6.0cm]{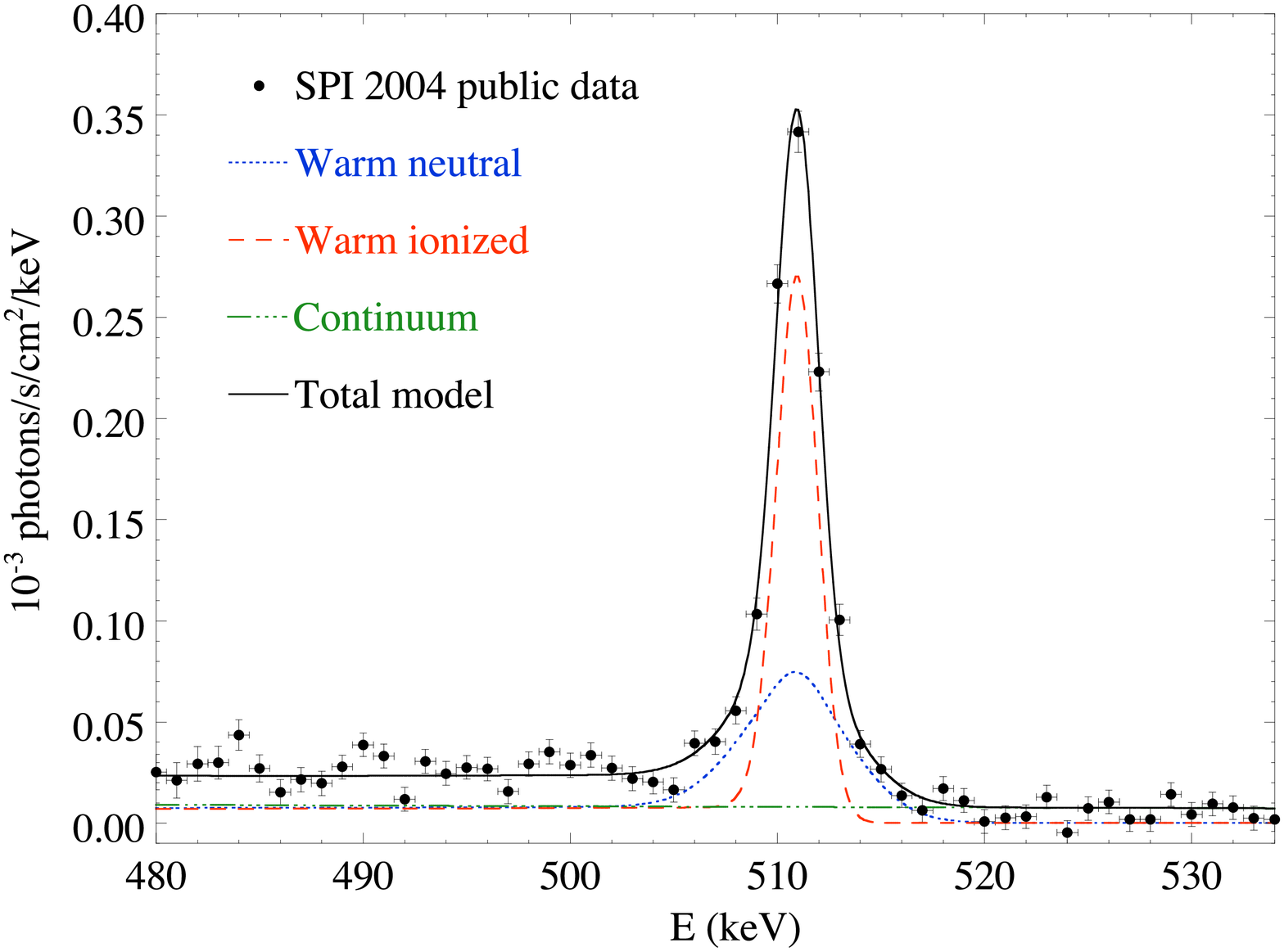}
\caption{Best fit of the spectrum measured by SPI with the warm components 
of the ISM and the Galactic continuum. Contributions from the molecular, 
cold and hot components are not needed to explain the data.
\label{fig:spifit2}}
\end{figure}

The parameters of this model are the phase fractions ($f_{i}$), the 
grain fraction ($x_{gr}$), the annihilation flux ($I_{e^{+}e^{-}}$) 
and the amplitude of the Galactic diffuse continuum 
at 511~keV ($A_{c}$; see Eq.~\ref{eq:spism}). 
Figure \ref{fig:spifit2} shows the best-fit model and Table 
\ref{tab:spifit2} the corresponding parameters. The 1$\sigma$ 
statistical uncertainties in the phase fractions were calculated 
separately by searching for the 68.3\%\ confidence interval of a 
phase fraction allowing the other parameters to vary but keeping 
the constraints $\Sigma_{i} f_{i}$ = 1. 
The total annihilation flux and the continuum amplitude at 511 keV 
obtained by this analysis are consistent with the independent model 
analysis. The spectral characteristics of the measured annihilation 
emission can be explained by positrons annihilating only in the warm 
ISM. Our best fit for the cold phase fraction is 0, however the 
upper-limit for this value is 23\%, so we cannot yet reject a 
significant contribution from this phase. The contribution of 
positrons in molecular clouds and hot gas is negligible since 
we obtain upper-limits of 8\%\ and 0.5\%, respectively. 

The total positronium fraction ($f_{Ps}$ = 93.5$^{+0.3}_{-1.6}$ \%) is 
calculated according to the intrinsic positronium fraction of each 
phase predicted by the GJG05 model, weighted by the phase fraction 
$f_{i}$. The positronium fraction in each phase is the sum of the 
contributions of positroniums formed in flight and in thermal 
conditions (via charge exchange and radiative recombination). 
These contributions were calculated using the probabilities of 
charge exchange in flight and the annihilation rates tabulated 
in GJG05. It should be noted that in this analysis, the fraction 
of ortho-positronium continuum is tied to the phase fractions.

Applying the method described in section \ref{s31}, we find 9$\%$ of 
systematic errors with this analysis, leading to the systematic 
uncertainties presented in Table \ref{tab:spifit2} for the molecular 
and cold phase fractions, while we obtained $^{+0.00}_{-0.04}$ of 
systematic uncertainty for the warm neutral phase fraction. However,
when the fit of the phase fraction is performed with a degradation parameter 
($\epsilon_{d}$) of 0.2 keV (0.4 keV), the optimal fractions are
51$\%$ (47$\%$) and 49$\%$ (53$\%$) for the warm neutral and warm 
ionized phases, respectively, the other fractions do not change and 
the fraction of positronium is 0.938 (0.933). The uncertainty in the 
slope of the continuum power-law model does not affect the results of 
the fit. Then to be conservative, the systematic errors for the warm 
neutral and ionized phase fractions are taken to be $^{+0.02}_{-0.04}$ 
and $\pm$0.02 respectively, and $\pm$0.3\%\ for the positronium fraction.

\begin{table}
     \caption{Measured ISM phase fractions obtained with 
     $\epsilon_{d}$ = 0.3 keV ($\chi^{2}$=176.4 for 148  
     degrees of freedom). The resulting positronium fraction is 
     0.935$^{+0.003}_{-0.016}$. The first set of error bars refers 
     to the 1$\sigma$ statistical errors and the second set to the 
     systematic errors. 
     \label{tab:spifit2}}
     \begin{array}[b]{ll}
\noalign{\smallskip}
\hline
\hline
\noalign{\smallskip}
\mbox{Parameters}  & \mbox{Measured values} \\
\hline
\noalign{\smallskip}
\mbox{f$_{m}$ (Molecular)} & 0.00 \; ^{+0.08}_{-0.00} \; ^{+0.02}_{-0.00} \\
\noalign{\smallskip}
\mbox{f$_{c}$ (Cold)} & 0.00 \; ^{+0.23}_{-0.00} \; ^{+0.04}_{-0.00} \\
\noalign{\smallskip}
\mbox{f$_{wn}$ (Warm Neutral)} & 0.49 \; ^{+0.02}_{-0.23} \; ^{+0.02}_{-0.04} \\
\noalign{\smallskip}
\mbox{f$_{wi}$ (Warm Ionized)} & 0.51 \; ^{+0.03}_{-0.02} \; ^{+0.02}_{-0.02} \\
\noalign{\smallskip}
\mbox{f$_{h}$ (Hot)} & 0.00 \; ^{+0.005}_{-0.00} \; ^{+0.00}_{-0.00} \\
\noalign{\smallskip}
\mbox{x$_{gr}$ (Grain fraction)} & 0.00 \; ^{+1.20}_{-0.00} \; ^{+0.20}_{-0.00} \\
\noalign{\smallskip}
  \hline
\end{array}
\end{table}

%

We now present the salient 
characteristics of the different spectral forms of the 
annihilation radiation that emerges from the different 
phases of the ISM; this will serve as a guide in the 
determination of the relative contributions of the 
different phases. The annihilation spectra for the different 
ISM phases can be characterized by the sum of three components: 
a narrow line due to the annihilation of thermalized positrons 
(except in the hot phase where the line is broad: a width of 
11 keV), a broad line emitted by the annihilation of positronium 
formed in flight, and an ortho-positronium continuum. 

Table \ref{tab:spchar} summarizes the spectral characteristics 
of the annihilation emission in the different phases. The widths 
of the 511 keV line listed in this table were extracted from Table 
3 of GJG05. The relative intensities of the different components 
($R_{i}$) and the fraction of positronium ($f_{Ps}$) for each phase 
were calculated using the fractions of positronium formed in flight 
in each case and the annihilation rates tabulated in GJG05 (see 
Tables 2 \&\ 4 therein), assuming $x_{gr}$ = 0. Since annihilation 
in dust grains is significant only in the hot phase (see section 2.2), 
we also calculated the spectral characteristics ($\Gamma_{i}$, 
$R_{i}$ \& $f_{Ps}$) of the annihilation emission from this phase 
with $x_{gr}$ = 1. The spectral characteristics obtained by the 
best-fit models are also shown for comparison. They were calculated 
using the results presented in Tables \ref{tab:spifit1} and 
\ref{tab:spifit2} of the present paper. By comparing the relative 
intensities $R_{i}$ of the annihilation features in each phase with 
the measured ones, we can conclude that the annihilation 
emission from a single phase cannot explain the measured spectral 
characteristics. A combination of annihilation emission from several 
phases is needed.

In the warm ionized phase, 
$\approx$87.4\% of positrons annihilate after thermalization, 
forming positroniums by radiative recombinaison with free electrons, 
while the remaining positrons annihilate directly with free electrons. 
In both annihilation processes, the width of the 511 keV line (0.98 
keV) can explain the narrow line component of the measured spectrum 
(Fig. \ref{fig:spifit1} and \ref{fig:spifit2}).
There is no formation of positronium in flight in this phase 
since the totality of H is ionized, thus no broad line is emitted. 
In the warm neutral phase, 94\% of positrons 
form positroniums in flight; one-fourth of them annihilate 
producing a 511 keV line emission with the 5.8 keV width required 
to explain the broad line component observed in the SPI data. 
The remaining $\approx$6\% of positrons thermalize and have sufficient 
kinetic energy to form positroniums by charge exchange with H atoms 
-- one-fourth of these positroniums annihilate producing a narrow 
($\Gamma_{n}$ = 1.16 keV) but weak ($R_{n} \sim$ 1.2 \%) 511 keV 
line. Consequently, the total positronium fraction in the warm neutral
phase is $\approx$99.9\%. 

Combining 49\% of annihilation in the warm 
neutral phase with 51\% of annihilation in the warm ionized phase as 
derived by the best-fit of the phase fractions yields $\approx$94\% 
total positronium fraction and the measured relative fluxes $R_{i}$. 
The broad line component with the 
required width can also be due to the annihilation of positroniums 
formed in flight in cold gas. However, most of the thermalized positrons 
annihilate directly with H in this phase (the temperature is too low 
in the cold gas to allow the formation of positronium by charge 
exchange), and therefore the total fraction of positronium is 
only due to the positroniums formed in flight ($\approx$94\%). 
Consequently, annihilation emission from the cold gas alone, 
or even combined with the annihilation emission from the warm 
ionized phase cannot (1) reproduce the measured positronium 
fraction and (2) explain the shape of the annihilation spectrum. 
However, a mix of annihilation emission from the cold gas and the 
warm neutral phase both combined with $\approx$51\% of annihilation 
emission from the warm ionized phase can reproduce the measured spectrum. 
This explains the uncertainty values on the cold and 
warm neutral phase fractions obtained by the best fit (see Table 
\ref{tab:spifit2}). Similar conclusions hold for the contribution of 
the annihilation emission from the molecular medium, which in addition 
is characterized by (1) narrow and broad 511 keV line widths that are 
both slightly larger than the cold phase's, and (2) a positronium 
fraction lower than those of the cold gas and the warm phase 
(see Table \ref{tab:spchar}). These differences significantly reduce 
the possible contribution of the molecular medium to the best-fit model. 
For the hot phase, the widths of the calculated annihilation line, 
with or without grains, are too broad and the positronium fraction too low 
for this phase to contribute substantially in the model.

\begin{table}
     \caption{Spectral characteristics of the annihilation radiation 
     in the different phases. $\Gamma_{n}$ and $\Gamma_{b}$ are the 
     width (FHWM) of the narrow and broad annihilation line, 
     respectively. $R_{n}$, $R_{b}$ and $R_{3\gamma}$ are the relative 
     flux of the narrow line, broad line and ortho-positronium, 
     respectively. $f_{Ps}$ is the fraction of positronium. Values 
     of $R_{i}$ and $f_{Ps}$ in the different phases were calculated 
     using the fraction of positronium formed in flight and the 
     annihilation rates tabulated in GJG05. The measured spectral 
     characteristics are shown for comparison. They were calculated 
     using the best-fit values of the free parameters of
     the ``Independent model'' (Table \ref{tab:spifit1}) 
     and the ``ISM model'' (Table \ref{tab:spifit2}).
     \label{tab:spchar}}
     \begin{array}[b]{lcccccc}
\noalign{\smallskip}
\hline
\hline
\noalign{\smallskip}
\mbox{Phase} & \Gamma_{n} & \Gamma_{b} & R_{n} & R_{b} & R_{3\gamma} & f_{Ps} \\
 & \mbox{(keV)} & \mbox{(keV)} & \mbox{(\%)} & \mbox{(\%)} & \mbox{(\%)} & \mbox{(\%)} \\
\hline
\noalign{\smallskip}
\mbox{Molecular} & 1.71 & 6.4 & 8.4 & 16.7 & 74.9 & 88.8 \\
\noalign{\smallskip}
\mbox{Cold} & 1.56 & 5.8 & 4.3 & 17.4 & 78.3 & 94.1 \\
\noalign{\smallskip}
\mbox{Warm Neutral} & 1.16 & 5.8 & 1.2 & 17.1 & 81.7 & 99.9 \\
\noalign{\smallskip}
\mbox{Warm Ionized} & 0.98 & - & 25.9 & 0.0 & 74.1 & 87.4 \\
\noalign{\smallskip}
\mbox{Hot (x$_{gr}$=0)} & - & 11.0 & 0.0 & 59.2 & 40.8 & 41.9 \\
\noalign{\smallskip}
\mbox{Hot (x$_{gr}$=1)} & 2.0 & 11.0 & 48.9  & 5.3 & 45.8 & 17.7 \\
\noalign{\smallskip}
\hline
\noalign{\smallskip}
\mbox{Independent model} & 1.3 & 5.4 & 13.8 & 8.0 & 77.8 & 96.7 \\
\noalign{\smallskip}
\mbox{ISM model} & 0.98 & 5.8 & 13.8 & 8.4 & 77.8 & 93.5 \\
\noalign{\smallskip}
  \hline
\end{array}
\end{table}


\section{\label{s4}Discussion}

Our analysis suggests that Galactic positrons annihilate primarily
in the warm phases of the ISM. A similar conclusion was reported 
in previous analyses (Harris et al. 1998; Churazov et al. 2005). 
Since positron annihilation takes primarily place in the Galactic 
bulge region (Kn\"odlseder et al. 2005), we now compare the current 
knowledge about the ISM in this area (i.e., within $\sim$ 600 pc 
radius of the Galactic Centre, corresponding to an approximative 
bulge size of 8$^{o}$ FWHM) with our results.

\subsection{Gas content in the Galactic bulge}

The gas content in the Galactic center region is not well known. 
The gas content of the Galactic bulge as well 
as its influence on the morphological and spectral characteristics of 
annihilation emission are under study and will be presented in 
a future paper (Gillard et al. in preparation). 
Launhardt et al. (2002) analysed IRAS and COBE 
data and showed that the nuclear bulge (region inside a Galactocentric
radius of $\approx$ 230 pc with a scale height of $\approx$ 45 pc) 
contains 7 $\times$ 10$^{7}$ M$_{\odot}$ of hydrogen gas (the mass of 
2 $\times$ 10$^{7}$ M$_{\odot}$ quoted in their paper has to be 
corrected by a factor 3.5; Launhardt, private communication). 

Launhardt et al. (2002) also argued that roughly 90\%\ of the interstellar 
mass in this region is trapped in small high-density ($\sim$10$^{4}$ cm$^{-3}$) 
molecular clouds with a volume filling factor of a few \%, while the 
remaining $\sim$10\% is homogeneously distributed and constitutes the 
intercloud medium with an average density $\sim$10 cm$^{-3}$ and 
probably a high ionization fraction.
From radio observations, Mezger \&\ Paul (1979) deduced a mass of HII 
gas in this region of 1.4 $\times$ 10$^{6}$ M$_{\odot}$ with an 
electron density n$_{e} \sim$~10 cm$^{-3}$ and an electron temperature 
T$_{e} \sim$~5000 K. Observations of the 21 cm line yield a HI mass of 
3.1 $\times$ 10$^{6}$ M$_{\odot}$ in this region (Rohlfs \&\ Braunsfurth 1982). 
The mass of H$_{2}$ in the nuclear bulge is estimated by subtracting 
this HI mass and the HII mass derived from the model of Lazio \&\ 
Cordes (2003)\footnote{The mass of HII derived from the model of Lazio \&\ 
Cordes (2003) is in agreement with the measurements performed by Mezger \&\ Paul 
(1979)} from the total mass of hydrogen gas measured by Launhardt et al. 
(2002). 

The rest of the gas in the Galactic bulge is contained in the asymmetric 
Galactic bar. Its shape can be approximated by an ellipsoid with 
semi-major axis $\sim$ 1.75~kpc and semi-minor axes $\sim$ 0.6~kpc
(Bissantz, Englmaier \&\ Gerhard 2003). The interstellar gas is not uniformly 
distributed across the bar, which makes its mass difficult to estimate.
Mezger et al. (1996) derived a mass of HI gas 
$\sim$~4~$\times$~10$^{7}$~M$_{\odot}$ in the bar from 
far infra-red and C$^{18}$O line observations. We can resonably assume 
that the HI gas mass is equally distributed between cold and warm 
neutral gases, as in the Galactic disk (Ferri\`ere 1998). Based on 
$^{13}$CO observations (Combes 1991) and a recent value of the abundance 
ratio $^{13}$CO/H$_{2}$ (Martin et al. 2004) we estimate a mass of 
H$_{2}$ gas $\sim$ 2 $\times$ 10$^{7}$ M$_{\odot}$ in the bar.

For the warm and hot ionized phases we consider the volume inside 600 
pc for which the Galactic free-electron density model of Lazio \&\ Cordes 
(2003) yields an HII gas mass of $\sim$ 2 $\times$ 10$^{6}$ M$_{\odot}$. 
We assume that 90$\%$ of this mass is in the warm ionized phase and 10$\%$ 
in the hot phase, similar to the proportion found in the Galactic disk 
(Ferri\`ere 1998).

From this mass model and the rough geometrical distribution of the gas 
in the Galactic bulge region, we derive the space-average densities 
$\langle n_{i} \rangle$ of the five interstellar phases. Since gravity 
is stronger in the Galactic bulge than near the Sun, we expect the gas 
to be more compressed there than locally. 
We then estimate the true density $n_{i}$ of each phase in the bulge by 
multiplying its true density $n_{i,\odot}$ measured near the Sun\footnote{with 
$n_{i,\odot}$=1000, 40, 0.4, 0.21 and 3.4$\times$10$^{-3}$ cm$^{-3}$ in 
the molecular, cold, warm neutral, warm ionized and hot phase 
respectively (Ferri\`ere 1998).}, by a common ``compression factor'' 
$f_{c}$, whose value is set by the requirement that $\sum_{i} 
\Phi_{i}$ = 1, where $\Phi_{i}$=$\langle n_{i} \rangle / n_{i}$ is the 
volume filling factor of phase i. This requirement leads to 
$f_{c}=\sum_{i} \langle n_{i} \rangle / n_{i,\odot} \approx$~3.6, and 
hence to the densities and filling factors listed in Table \ref{tab:bulge}. 
The true density obtained 
for the molecular gas with this method is in agreement with the 
observations of Launhardt et al. (2002), Martin et al. (2004) and 
Stark et al. (2004).

\begin{table}
     \caption{Estimates for the Galactic bulge parameters and 
     resulting consequence on the lifetime and diffusion of 1 MeV positrons 
     (see text). $\langle n \rangle$ and $n$ are the space-averaged and 
     true densities, respectively. $\Phi$ is the volume filling 
     factor. $l/2$ is the typical half-size of the 
     phase. K$_{ql}$ is the energy threshold above which the quasilinear 
     diffusion is valid. $\Delta t_{ql}$ and $d_{ql}$ are the time taken 
     and the distance travelled by positrons in quasilinear diffusion 
     regime. $\Delta t_{coll}$ and $d_{coll}$ are the time taken 
     and the maximum distance travelled by positrons in the collisional diffusion 
     regime. $d_{ann}$ is the maximum distance travelled by 
     thermalized positrons. $\tau_{ann}$ is the lifetime of 
     thermalized positrons. $d_{max}$ is the total distance travelled 
     by positrons ($d_{max} = d_{ql}+d_{coll}+d_{ann}$). The lifetime 
     for thermalized positrons in the hot phase was calculated assuming a 
     normal abundance of interstellar grains (x$_{gr}$=1). \label{tab:bulge}}
     \begin{array}[b]{lccccc}
\noalign{\smallskip}
\hline
\hline
\noalign{\smallskip}
\mbox{}  & \mbox{molecular}& \mbox{cold}& \mbox{warm}& \mbox{warm}& \mbox{hot} \\
\mbox{}  & \mbox{} & \mbox{} & \mbox{neutral}& \mbox{ionized}& \mbox{} \\
\hline
\noalign{\smallskip}
\langle n \rangle \mbox{(cm$^{-3}$)} & \mbox{1.58}& \mbox{0.26}& \mbox{0.26}& \mbox{0.08}& \mbox{8.9 $\times$10$^{-3}$} \\
\noalign{\smallskip}
n \mbox{(cm$^{-3}$)} & \mbox{3600}& \mbox{146}& \mbox{1.46}& \mbox{0.77}& \mbox{0.012} \\
\noalign{\smallskip}
\Phi & \mbox{0.0004}& \mbox{0.002}& \mbox{0.18}& \mbox{0.10}& \mbox{0.72} \\
\noalign{\smallskip}
l/2 \mbox{(pc)} & \mbox{3-30}& \mbox{$\sim$5}& \mbox{0.1-50}& \mbox{10-100}& \mbox{50-100} \\
\noalign{\smallskip}
\hline
\noalign{\smallskip}
K_{ql} \mbox{(keV)} & \mbox{10$^{-3}$}& \mbox{0.03}& \mbox{2.9}& \mbox{5.5}& \mbox{270} \\
\noalign{\smallskip}
\Delta t_{ql} \mbox{(yr)} & \mbox{39}& \mbox{10$^{3}$}& \mbox{10$^{5}$}& \mbox{10$^{5}$}& \mbox{2.7 $\times$ 10$^{6}$} \\
\noalign{\smallskip}
d_{ql} \mbox{(pc)} & \mbox{1.0}& \mbox{4.8}& \mbox{47.8}& \mbox{43.9}& \mbox{264} \\
\noalign{\smallskip}
\hline
\noalign{\smallskip}
\Delta t_{coll} \mbox{(yr)} & \sim \mbox{0} & \sim \mbox{0}& \mbox{31}& \mbox{56}& \mbox{6.6 $\times$10$^{5}$} \\
\noalign{\smallskip}
d_{coll} \mbox{(pc)} & \sim \mbox{0} & \sim \mbox{0} & \mbox{0.10}& \mbox{0.09}& \mbox{5210} \\
\noalign{\smallskip}
\hline
\noalign{\smallskip}
\tau_{ann} \mbox{(yr)} & \mbox{22}& \mbox{3500}& \mbox{1.3 10$^{4}$}& \mbox{3.4 10$^{4}$}& \mbox{9.4 10$^{6}$} \\
\noalign{\smallskip}
d_{ann} \mbox{(pc)} & \sim \mbox{0} & \sim \mbox{0} & \mbox{0.04}& \mbox{0.004}& \mbox{172} \\
\noalign{\smallskip}
\hline
\noalign{\smallskip}
d_{max} \mbox{(pc)} & \mbox{1.0} & \mbox{4.8} & \mbox{47.9}& \mbox{44.0}& \mbox{5.6 $\times$ 10$^{3}$} \\
\noalign{\smallskip}
\hline
\end{array}
\end{table}

If positrons are generated uniformly in the Galactic bulge and annihilate 
in situ (i.e. without propagation) then, in stationary conditions, 
the phase fraction f$_{i}$ of each phase must be equal to its filling 
factor $\Phi_{i}$. 
In this case, one would expect the hot medium to be the dominant 
component. However, in a 0.01~cm$^{-3}$ density hot medium, 1~MeV
positrons thermalize in $\sim$4~$\times$10$^{6}$ years and then 
annihilate on a timescale $\sim$10$^{8}$ years if there are no 
interstellar grains in this phase; in the standard grain model 
the annihilation timescale is $\sim$10$^{7}$ years (GJG05). 
With such long timescales it is likely that positrons escape the hot
medium (before they can annihilate), either by propagation 
or following an encounter with a supernova shock wave (timescales $\sim$ 
0.5--1 Myr; Cox 1990). Positrons escaping the hot medium then have 
a high probability of entering a warm phase due to the large filling 
factor of this phase (c.f. Table \ref{tab:bulge}). If such positrons 
annihilate in the warm phases, then the phase 
fractions between the warm neutral and warm ionized phase should be 
close to the relative values of their filling factors. In order to 
verify whether positrons escape a given ISM phase or not, it 
is necessary to estimate the distance they travel within this phase 
and compare it with the typical half-size of the phase. 

\subsection{Propagation of positrons}

The distance travelled by positrons depends on their initial velocity 
and their energy loss rate, which is a function of the density and 
the ionization fraction of the ambient medium. 
The timescale for 1 MeV positrons to thermalize in a warm medium is 
$\tau_{w} \approx$ 10$^{5}$ years. Positrons then annihilate with a 
timescale $\approx$3.4$\times$10$^{4}$ years in the 0.8 cm$^{-3}$ density warm 
ionized phase. In the 1.5 cm$^{-3}$ density warm neutral gas, positrons 
that do not form a positronium in flight by charge exchange with H, take 
$\approx$1.3$\times$10$^{4}$ years to annihilate. Typical sizes of warm and hot 
regions are $l_{w} \sim l_{h} \sim$100 pc. Roughly speaking, positrons 
will escape the hot phase if the distance travelled by diffusion in a time 
$\tau_{h} \sim$10$^{7}$ years (slowing down time plus annihilation 
time in the standard grain model) is greater than the typical 
half-size of hot region, $l_{h}/2 \sim$ 50 pc, i.e., 
if the diffusion coefficient $D$ is greater than $D_{min} = 
\frac{l_{h}^{2}}{24 \tau_{h}} \sim$1.3$\times$10$^{25}$ cm$^{2}$s$^{-1}$. 
On the other hand, positrons that enter the warm 
phases will annihilate if the distance travelled by diffusion in a 
time $\tau_{w} \sim$ 10$^{5}$ years is less than $l_{w}/2 \sim$ 
50 pc, i.e., if $D$ is smaller than $D_{max} = \frac{l_{w}^{2}}{24 
\tau_{w}} \sim$1.3$\times$10$^{27}$ cm$^{2}$s$^{-1}$.
These limits have to be compared with the quasilinear diffusion 
coefficient (Melrose 1980) which can be expressed as:

\begin{equation}
   D_{ql}(E) = D_{B} \left(\frac{r_{L}}{\lambda_{max}}\right)^{1-\delta} 
   \eta^{-1}
     \label{eq:dql}
\end{equation}
\noindent
with $D_{B}=\frac{1}{3} r_{L} v$ the Bohm diffusion coefficient, 
$r_{L}$ the positron gyroradius, $v$ the positron velocity, 
$\lambda_{max}$ the maximum scale of the turbulence, $\delta$ = 5/3  
for a Kolmogorov turbulent spectrum, and 
$\eta$ = $\delta$B$^{2}/ \langle$B$\rangle^{2}$ the 
relative perturbation in magnetic field pressure which is often 
approximated to 1, since the turbulent component of the ISM magnetic 
field has been estimated to be of the same order of magnitude as 
the regular magnetic field. 

The magnetic field strength in the Galactic Centre region was estimated 
to be $\sim$10 $\mu$G (Sofue et al. 1987; La Rosa et al. 2005). 
The maximum scale $\lambda_{max}$ was estimated to be $\sim$ 100 pc 
from measurements of ISM turbulence (Armstrong et al. 1995). Then for 
1~MeV positrons in the Galactic bulge, $D_{ql} \sim$ 3$\times
$10$^{26}$ cm$^{2}$s$^{-1}$. This coefficient is well above 
$D_{min}$ and below $D_{max}$ and leads support to our hypothesis that 
positrons indeed escape the hot phase and subsequently annihilate in the 
warm phases. 

However, quasilinear diffusion is valid only when positrons 
are in resonance with Alfv\'en waves. This condition is satisfied 
when:

\begin{equation}
   \gamma \beta > \frac{m_{p}v_{A}}{m_{e}c} = 12.85  \times 10^{-3} 
   \frac{B_{\mu G}}{\sqrt{n_{cm^{-3}}}}
    \label{eq:gambet}
\end{equation}
\noindent
with $\gamma$ the Lorentz factor, $\beta=\frac{v}{c}$, 
$m_{p}$ the proton mass, $m_{e}$ the positron mass and $v_{A}$ 
the Alfv\'en speed.

When the kinetic energy of positrons drops below a 
given threshold $K_{ql}$, quasilinear diffusion theory breaks 
down and the diffusion regime changes. $K_{ql}$ is calculated for each 
phase using Eq. (\ref{eq:gambet}) together with B = 10$\mu$G and the densities 
given in Table \ref{tab:bulge}. 
The slowing down times $\Delta t_{ql}$ for 1 MeV positrons to reach 
$K_{ql}$ are listed in Table \ref{tab:bulge} as well as 
the associated distances $d_{ql}$ which are obtained by:

\begin{equation}
   d_{ql} = \sqrt{ \int_{K_{ql}}^{1MeV} 6 D_{ql}(E) 
   \left(\frac{dE}{dt}\right)^{-1} dE }
    \label{eq:distql}
\end{equation}

\noindent
The distances $d_{ql}$ are lower than or of the same order as the typical 
half-sizes of the corresponding phases (see Table \ref{tab:bulge}) 
except for the hot gas. 

The diffusion regime of positrons with kinetic energies below $K_{ql}$ 
is uncertain (interstellar winds, resonance with other plasma waves\ldots) and 
is currently under study (Marcowith et al., in preparation). 
However we can estimate an {\it upper-limit} to the distance $d_{coll}$ travelled 
by these positrons assuming that they propagate in a {\it collisional} regime. 
In this case, the distance is given by:

\begin{equation}
   d_{coll} = \sqrt{ \int_{E_{lim}}^{K_{ql}} 6 D_{coll}(E) 
   \left(\frac{dE}{dt}\right)^{-1} dE }
    \label{eq:distcol}
\end{equation}

\noindent
where $E_{lim}$ is the lowest kinetic energy for which 
positrons are able to form a positronium in flight by charge 
exchange in the molecular, cold and warm neutral phases or are 
thermalized ($E_{lim}$ = $\frac{3}{2}$k$T$) 
in the warm ionized and hot phases (charge exchange does not happen in these 
phases since all the hydrogen is ionized). $\frac{dE}{dt}$ is the energy-loss 
rate. Accounting for the streaming of positrons in the ISM, at a 
characteristic velocity $\sim v_{A}$, we write the diffusion 
coefficient $D_{coll}(E) = v_{A} \times \lambda_{coll}(E)$ where $\lambda_{coll}(E)$ 
is the distance over which interactions gradually deflect positrons with 
kinetic energy $E$ by 90$^{o}$. $\lambda_{coll}$ was calculated according 
to Lang (1974) for a fully ionized plasma and using the approach 
of Emslie (1978) for neutral hydrogen. When positrons are thermalized, 
the distance $d_{ann}$ they travel before annihilation is given by:

\begin{equation}
   d_{ann} = \sqrt{6 D_{coll} \tau_{ann}}
    \label{eq:distth}
\end{equation}

\noindent
where $\tau_{ann}$ is the annihilation lifetime of thermalized positrons. 
The resulting values of $d_{max}=d_{ql}+d_{coll}+d_{ann}$ (see Table 
\ref{tab:bulge}) confirm that 1~MeV positrons injected in the hot 
phase should escape it, while those injected in other phases should 
annihilate in them. Since positrons leaving the hot phase most likely
enter the warm phases (due to their large filling factors) we expect 
that positron annihilation occurs mostly in the warm phases.

It has to be noted that several authors (see Morris \&\ Serabyn 1996 
and references therein) estimate magnetic field values of 
$\sim$1mG in the Galactic centre region. This is 2 orders of magnitude 
larger than the value (10$\mu$G) used in our study. If we assume that 
this value is effective everywhere in the Galactic bulge, then this 
increases $v_{A}$ by 2 orders of magnitude and rules out quasilinear 
diffusion in the warm and hot phases for 1~MeV positrons (see Eq. 
\ref{eq:gambet}). Consequently, we can only derive upper limits to 
the distances travelled by positrons in these phases, by assuming that 
they propagate in a collisional regime. Here again, a detailed investigation 
of the diffusion of positrons in a non-quasilinear regime is 
required to derive relevant distances of propagation.

\subsection{Initial kinetic energy of positrons}

\begin{figure}[tb]
\centering
\includegraphics[width=8.8cm,height=6.0cm]{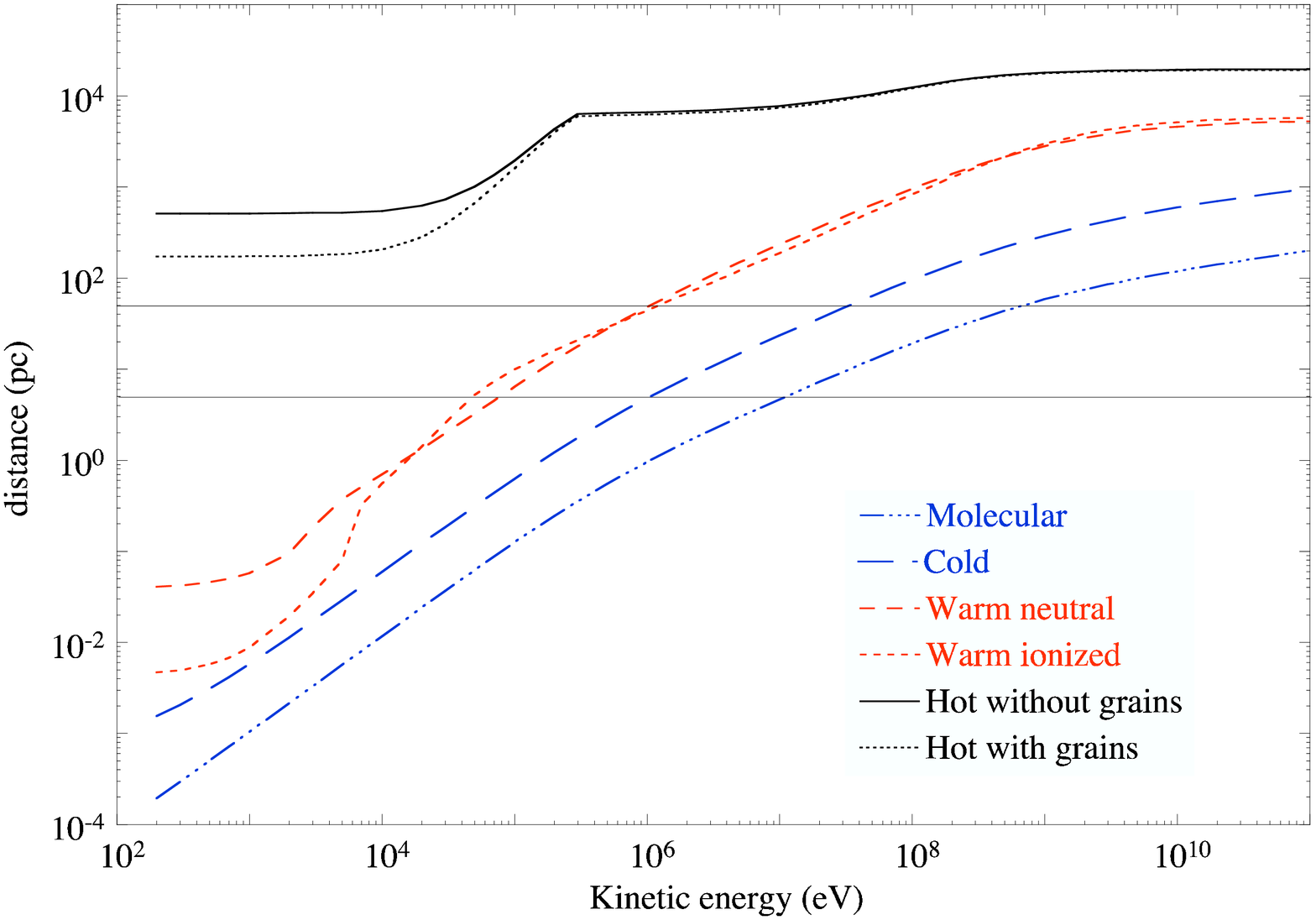}
\caption{Maximum distance travelled by positrons as a function of their 
kinetic energy. Typical half-sizes of warm and hot regions ($\sim$50 pc) and 
typical sizes of cold and molecular regions ($\sim$5 pc) are shown 
for comparison.
\label{fig:dist}}
\end{figure}

The above estimates on the transport of positrons 
are based on the assumption that their initial kinetic energy is 
1 MeV, which is a typical value for positrons emitted by radioactive nuclei. 
The actual mean energy of positrons from $^{56}$Co, one of the proposed 
candidates for the source of Galactic positrons, is 0.6~MeV. 
Figure \ref{fig:dist} shows the maximum distance 
$d_{max}$ = $d_{ql}$+$d_{coll}$+$d_{ann}$ travelled by positrons before 
annihilating as a function of their initial kinetic energy in a 10 
$\mu$G magnetic field. The conclusions 
derived above, assuming 1~MeV positrons, obviously hold for positrons 
with E$<$1 MeV even for the hot phase. Positrons with 
initial kinetic energy above $\sim$1~MeV would escape the warm 
and hot phases and would then have a chance to annihilate in molecular 
clouds provided they do not escape the Galactic bulge. However, the 1~MeV 
limit should not be considered as very constraining, since it was 
derived using rough estimates for the size of each phase (see 
Table \ref{tab:bulge}). Moreover, we have to keep in mind that 
the calculated distances $d_{max}$ quantify the spatial extent 
of the distribution of annihilating positrons around the source. 
Then the fraction of positrons produced in a warm phase with initial 
energy of e.g. 2~MeV that escape this warm phase to annihilate in cold 
or molecular clouds, would probably be sufficiently low to be in 
agreement with the phase fractions derived by the spectral analysis, since
most of these positrons would annihilate in the warm phase.

In our discussion, we have assumed that positrons are generated uniformly 
in the Galactic bulge. This assumption is valid in the case of 
numerous positron sources, including light dark matter annihilating into 
electron-positron pairs (Boehm et al. 2003). However, such positrons should 
have an initial kinetic energy lower than a few MeV to explain the 
measured spectrum. Other possible uniformly distributed sources in the 
Galactic bulge are type Ia supernovae, novae and low mass X-ray binaries, 
all of which belong to the old stellar population. However, these sources 
produce their own warm and hot phases by heating the surrounding ISM, so 
that they do not inject positrons in cold atomic and molecular regions.

Low energy ($<$1MeV) positrons produced in the warm phase probably stay in 
this phase and annihilate at a distance $\la$50pc from the sources. 
When such positrons are produced in the hot medium, they are likely to 
leave this medium and enter the warm phase within which they 
will annihilate. Then, assuming that the typical size of the hot phase 
does not exceed $\sim$200~pc and that positrons are produced at the 
edge of this phase, then the maximum distance covered by such positrons 
is $\la$250 pc from the source. Therefore a single source releasing 
positrons in a warm or hot region (e.g. a gamma-ray burst as suggested 
by Parizot et al. 2005) might have difficulty accounting for the 
observed spatial extent of the annihilation emission, which covers a 
radius $\sim$600 pc around the Galactic centre. 

\section{\label{s5} Conclusions}

The positron-electron annihilation emission spectrum can be explained 
by narrow and broad 511 keV lines plus an ortho-positronium 
continuum. The detection significance of the broad line is $\approx$3.2$\sigma$. 
The broad line width of (5.4$\pm$1.2) keV FWHM is in agreement with 
the value calculated by GJG05 for positronium formed in flight 
by charge exchange with H ($\approx$5.8 keV). This value is also in agreement 
with the width ($\approx$5.3 keV) calculated by Churazov et al. (2005). 
The narrow line width is (1.3$\pm$0.4) keV FWHM. This width is consistent 
with the $\approx$1 keV width of positrons annihilating by radiative 
recombination in the warm ionized medium.

Galactic positrons seem to annihilate mostly in the warm phases of the 
ISM. These results are in agreement with conclusions of Harris et 
al. (1998) and Churazov et al. (2005). We estimate that $\approx$50 
\% of the annihilation emission comes from the warm neutral phase and 
$\approx$50 \% from the warm ionized phase. The contribution of 
molecular clouds and the hot phase are less than 8\% and 0.5\%, respectively.
We cannot exclude from our spectral analysis that a significant 
fraction ($<$23\%) of the emission comes from cold gas. However 
in view of the gas content of the Galactic bulge, this fraction 
is expected to be negligible.

A preliminary study of the interstellar gas content and positron propagation
in the Galactic bulge shows that the phase fractions derived from the spectral 
analysis are in agreement with the relative filling factors of the warm 
and low temperature gases. The lack of detection of annihilation in 
molecular and cold atomic gases could be explained by their low filling 
factors. The spectral analysis suggests comparable 
amounts of annihilation in the warm neutral and warm ionized 
phases of the ISM. This is in good agreement with our expectation if 
the positron sources are uniformly distributed and if the initial kinetic 
energy of positrons is lower than a few MeV, otherwise they may escape 
the warm phase of the ISM and annihilate in molecular or cold atomic 
regions.

Despite its large filling factor, positrons 
do not annihilate in the hot gas because its density is low enough
to allow them to escape. Using quasilinear diffusion theory and 
assuming a magnetic field $\sim$10~$\mu$G in the Galactic bulge, we 
estimate that, down to low kinetic energies, positrons are generally 
confined in warm and cold regions. In the hot 
phase, they should be less confined and assuming that they are released 
in a collisional regime when their kinetic energy drops below 
$\sim$270~keV, 
they have sufficient kinetic energy to escape it. However, the diffusion 
regime of positrons with keV energies in a hot low-density plasma is not 
yet known. More detailed studies on the Galactic bulge gas content and on 
the diffusive regime of positrons as a function of their energy and 
the magnetic field strength are under way and will be presented in 
forthcoming papers.

According to the rough gas model described in section \ref{s4}, we 
expect the warm ionized component to dominate the annihilation 
emission in the nuclear bulge, while the warm neutral component should 
dominate in the Galactic bar. Additional exposure will 
make it possible to perform spectra in different regions of the Galactic 
Centre and confirm this prediction. 

Our understanding of Galactic positron physics can be improved by 
imaging and spectroscopic explorations of Galactic regions 
other than the central regions, particularly by measurements of the 
annihilation emission from the Galactic disk, which appears to be 
explainable by $^{26}$Al decay (Kn\"odlseder et al. 2005). 
While our understanding of positron annihilation in the Galactic Centre 
region is limited by the poor knowledge of both the sources and initial kinetic 
energy of positrons and the gas content in the Galactic bulge, 
the task would be easier for the disk emission, since: (1) we 
know that positrons from $^{26}$Al are released in the molecular ring 
region since COMPTEL onboard CGRO measured the spatial distribution 
of the 1.8 MeV gamma-ray line emitted during the radioactive decay of 
this isotope; (2) the average energy of such positrons has been 
measured to be 450 keV; 
(3) the distribution and characteristics of the gas in the molecular
ring region are known with better accuracy than in the Galactic bulge. 
Consequently, the measured spatial distribution of the
annihilation emission from the disk would teach us how far 
from the sources positrons annihilate and this will lead to estimates 
of their diffusion coefficient. The spectral characteristics of the disk 
emission would tell us in what phases of the ISM positrons annihilate. 
Since much $^{26}$Al is located in the molecular ring region, we expect to 
measure a strong 6.4~keV FWHM line component due to the annihilation of 
positronium formed in flight in H$_{2}$ (see Fig. 5 of GJG05).
On the other hand, the significance of the annihilation emission from 
the disk is still weak (3--4$\sigma$) after one year of the INTEGRAL 
mission. Additional exposure is needed to allow the spectral analysis 
to provide constraining results.

\bigskip
{\it Additional note: After acceptation of this manuscript, a related paper 
(Weidenspointner et al. 2005) recently submitted, confirms that 
the spatial distribution of the positronium continuum emission 
measured using SPI data, is consistent with the 511 keV line
emission distribution.}
\bigskip


\begin{acknowledgements}
We are grateful to R.J. Murphy for useful discussions on the energy 
loss rate of positrons in the ISM. We thank the anonymous referee 
for suggestions that have improved the quality of this paper.
\end{acknowledgements}



\end{document}